\documentclass{INTERSPEECH2023}

\usepackage{multirow}
\usepackage{tikz}
\usepackage{graphicx}
\usepackage{booktabs}
\usepackage{multirow}
\usepackage{tabularray}
\def\algbackskip{\hskip-\ALG@thistlm}
\usepackage{algorithm}
\usepackage{tabularray}
\usepackage{algpseudocode}
\newcommand {\MEL} {{\mathrm{M}}}
\newcommand {\R} {{\mathbb{R}}}

\newcommand {\om} {{\omega}}
\newcommand {\la} {{\lambda}}

\newcommand {\lau} {{\lambda_1}}
\newcommand {\lad} {{\lambda_2}}

\newcommand {\tS} {{\widetilde{S}}}

\newcommand{\ra}[1]{\renewcommand{\arraystretch}{#1}}

\interspeechcameraready 

\title{Wavelet Scattering Transform for Improving Generalization in Low-Resourced 
Spoken Language Identification}

\name{Spandan Dey, Premjeet Singh, Goutam Saha}

\address{Department of Electronics \& Electrical Communication Engineering,\\ Indian Institute of Technology Kharagpur, West Bengal, India}
\email{sd21@iitkgp.ac.in, premsingh@iitkgp.ac.in, gsaha@ece.iitkgp.ac.in}

\begin{document}

\maketitle

\begin{abstract}
Commonly used features in spoken language identification (LID), such as mel-spectrogram or MFCC, lose high-frequency information due to windowing. The loss further increases for longer temporal contexts. To improve generalization of the low-resourced LID systems, we investigate an alternate feature representation, wavelet scattering transform (WST), that compensates for the shortcomings. To our knowledge, WST is not explored earlier in LID tasks. We first optimize WST features for multiple South Asian LID corpora. We show that LID requires low octave resolution and frequency-scattering is not useful. Further, cross-corpora evaluations show that the optimal WST hyper-parameters depend on both train and test corpora. Hence, we develop fused ECAPA-TDNN based LID systems with different sets of WST hyper-parameters to improve generalization for unknown data. Compared to MFCC,  EER is reduced upto $14.05\%$ and $6.40\%$ for same-corpora and blind VoxLingua107 evaluations, respectively.
\end{abstract}


\section{Introduction}
\footnote{Accepted in INTERSPEECH 2023 (pre-camera-ready version)}In the context of spoken language identification (LID), the term    \emph{low-resourced} indicates the lack of large-scale multilingual speech corpora with verified ground truths. Developing effective low-resourced LID systems is important for  multilingual human-to-computer interaction applications for the global population~\cite{ambikairajah2011language}. Often, the generalization of the low-resourced LID systems is challenged due to training with a small in-house developed corpus, which lacks diversities in non-lingual characteristics~\cite{dey2021overview}. 
Towards this, we aim to improve the generalization of low-resourced LID systems by applying feature representations that are robust against non-lingual variations across multiple corpora. To assess generalization, we follow cross-corpora evaluation protocols, which are particularly useful for low-resourced scenarios~\cite{dey2021cross}.

Mel-spectrograms and mel frequency cepstral coefficients (MFCCs), are one of the most widely used features in the LID literature. Both these features were originally developed for automatic speech recognition (ASR). By applying  mel filterbanks on short-time Fourier transform (STFT) representations, these features generate features resembling the output of the cochlea~\cite{rabiner1993fundamentals}. The mel-based features further provide stability toward local time translation and deformation, usually up to a window of 25ms~\cite{anden2014deep}. This temporal span works well for phoneme recognition purposes in ASR but may not be the most suitable choice for recognizing languages, which requires longer phonotactic information as well~\cite{li2013spoken}. One of the key issues with these features is the loss of information due to the time-domain convolution with a low-pass filter due to windowing~\cite{anden2014deep}. For classification tasks requiring a larger temporal context, capturing these features over a longer temporal window incurs even more information loss.

Hence, in this work, we introduce alternative representations in the LID task, that provide stability against deformations and reduce information loss even with longer temporal contexts. Mallat proposed \emph{wavelet scattering transform} (WST) extending the MFCCs by computing modulation spectrum with the cascaded application of wavelet filter banks and modulus non-linearities~\cite{mallat2012group}. Abiding by the \emph{Lipschitz continuity} condition, WST extracts stable representations for variations due to time-shifts and time-warping deformations over a larger temporal span without losing the high-frequency information~\cite{anden2014deep}. 
In~\cite{anden2014deep}, the authors applied WST for music genre and phoneme classification. Environmental sound classification was performed utilizing WST in~\cite{bauge2013representing,kek2022intelligent,hajihashemi2022binaural}. Bruna et al.~\cite{bruna2013audio} used scattering moments to synthesize audio textures. Joy et al.~\cite{joy2020deep} applied scattering power spectrum for speech recognition. 
%Lostanlen et al.~\cite{lostanlen2021time} used joint time-frequency scattering method to cluster different instrument notes based on timbre similarity. A data-driven dimensionality reduction technique was applied over WST for playing technique recognition in~\cite{wang2022adaptive}. 
In~\cite{lostanlen2021time} and~\cite{wang2022adaptive}, WST was applied for music processing applications. Recently, the potentials of WST are explored in speech emotion recognition~\cite{singh2021deep} and speaker recognition tasks~\cite{ghezaiel2021hybrid}. 

However, to our knowledge, this is the first study applying scattering networks in a LID task. Therefore, we first systematically formulate the work, starting by exploring the fundamental questions:~\textbf{(Q1)} What are the WST hyper-parameters suitable for the LID tasks?~\textbf{(Q2)} Are the optimized hyper-parameters corpora dependent?~\textbf{(Q3)} How much performance improvements can we obtain from the conventional MFCCs by optimizing the WST hyper-parameters?~\textbf{(Q4)} Is scattering transform across the log-frequency dimension useful for LID?
After answering these fundamentals, we then focus on improving cross-corpora LID generalization with WST features. The answer to Q2 reveals the dependency of optimal WST hyper-parameters on both training and evaluation data. Hence, concerning unknown real-world test conditions, we develop multi-WST fusion based LID systems encompassing the representations generated with different WST hyper-parameters.

\section{Methodology}
\subsection{Shortcomings of Fourier-based representations}
Consider a signal $x(t)$ with its Fourier transform (FT) denoted by $\widehat{x}(\om)$. For a time-shift $c$ expressed by $x_c(t) = x(t-c)$, the corresponding FT is $\widehat{x}_c(\om) = e^{-i\om c}\,\widehat{x}(\om)$. The modulus of FT removes the additional phase part $e^{-i\om c}$ and makes the representation translation invariant, $|\widehat{x}_c(\omega)| = |\widehat{x}(\omega)|$. Let the short-time Fourier transform (STFT) or spectrogram of $x(t)$ is defined as $|\widehat{x}(t,\om)| = \left| \int x(u)\,\phi(u-t)\,e^{-i\om{}u}\,du\right|$. Here, $\phi$ is a window with duration $T$. For $c << T$, STFT is local time-shift invariance. However, STFT is not stable to time-warping deformations, which often takes place with audio data~\cite{anden2014deep}. Here, the notion of stability is defined by the Lipschitz continuity condition, which states that a transformed representation $\Phi(x)$ is stable to deformation by amount $\sup_t |\tau'(t)|$ if, 
\vspace{-.15cm}
\begin{equation}
\label{eq:lipschitz} 
||\Phi(x) - \Phi(x_\tau)|| \leq C\, \sup_t |\tau'(t)|\, ||x||~.
\vspace{-.35cm}
\end{equation} 

Here, $C > 0$ is a constant and measures the stability. Considering a deformation, $\tau(t) = \epsilon{}t$, where $0 < \epsilon \ll 1$ the Fourier transform of  $x_\tau(t)= x(t-\tau(t)) =  x((1-\epsilon)t)$, for $|\tau'(t)| < 1$, is
$\widehat{x}_\tau(\om) = (1-\epsilon)^{-1}\,\widehat{x}((1-\epsilon)^{-1}\om)$. Hence, time deformation leads to frequency translation of $\epsilon{}|\om|$. Following  Eq.~\ref{eq:lipschitz}, we obtain $||\,|\widehat{x}| - |\widehat{x_\tau}|\,|| \leq C\,\epsilon\,||x||$. This implies that Lipschitz continuity can be violated by spectrograms for higher values of $\om$.

To impose stability against time-warping deformation, averaging with mel filters ($\widehat \psi_{\la} (\om)$), with center frequency $\la$ is done over spectrograms, i.e.,  %$\MEL x(t,\la) = \frac{1}{2\pi} \int |\widehat{x}(t,\omega)|^2\,|\widehat{\psi}_{\la}(\om)|^2d\om~.$ 
\vspace{-.15cm}
\begin{equation}
	\label{eq:mfsc}
	\MEL x(t,\la) = \frac{1}{2\pi} \int |\widehat{x}(t,\omega)|^2\,|\widehat{\psi}_{\la}(\om)|^2d\om~,
 \vspace{-.25cm}
\end{equation}

where, $\widehat{x}(t,\om)$ is the FT of $x_t(u) = x(u)\phi(u-t)$. The time-domain equivalent of Eq.~\ref{eq:mfsc} is given as, 
\vspace{-.15cm}
\begin{equation}
\label{eq:mfsc_time_domain}
	\MEL x(t,\la) = \int \left|\int x(u)\phi(u-t)\psi_\la(v-u)du\right|^2\,dv
 \vspace{-.25cm}
\end{equation}

Now, if the length ($T$) of window $\phi$ is much larger than the temporal support of $\psi_{\la} (t)$, we can consider $\phi(t)$ as constant over the span of  $\psi_{\la} (t)$. Hence, considering $\phi(u-t)\psi_\la(v-u) \approx \phi(v-t)\psi_\la(v-u)$ in Eq.~\ref{eq:mfsc_time_domain}, 
\vspace{-.1cm}
\begin{align}
	\MEL x(t,\la) &\approx \int \left|\int x(u)\psi_\la(v-u)du\right|^2|\phi(v-t)|^2dv \\
\label{eq:mfsc_final}
	&= |x\star\psi_\la|^2\star|\phi|^2(t)~.
 \vspace{-.65cm}
\end{align}

The mel-filters $\widehat \psi_{\la} (\om)$ with center frequency $\la$ have constant-$Q$ bandwidths (BW) at high frequencies. So, the BW of $\widehat \psi_{\la} (\om)$ is $\la / Q$, which is sufficiently large at higher $\la$s to encompass stability to time-warping deformations. However, as evident from Eq.~\ref{eq:mfsc_final}, the windowing performed on the signal is equivalent to time averaging of spectrograms by the low-pass filter $\phi(t)$. Hence, there is an inherent downside in mel-spectrograms/MFCCs of high-frequency information loss. Instead of the standard window size, usually fixed to \SI{25}{\milli\second}, if some classification task demands a higher temporal context, the risk of higher information loss restricts the usefulness of mel-spectrograms/MFCCs. Hence, to restore the high-frequency information while maintaining stability to deformations over a longer span, wavelet scattering transform can be used.

\subsection{Wavelet scattering transform}
Scattering transform applies cascade of wavelet transforms, with constant Q-factor wavelet filters and modulus operators for restoring high-frequency information lost due to averaging. Consider, for $\la >0$ a dilated wavelet band pass filter $\psi_\la (t) = \la\, \psi(\la\, t)$ with frequency-domain representation $\widehat \psi_\la (\om) = \widehat \psi\Bigl(\frac{\omega} {\la}\Bigr)$. The center frequency of $\widehat \psi_\la (\om)$ is $\la$ (normalized) and BW is $\lambda/Q$, with $Q$ denoting the octave resolution of wavelet filters. Hence, $\la = 2^{k/Q}$ with $k\in \mathbb{Z}$. The wavelet filters span the entire frequency range of the input signal. Each filter $ \psi_\la (t)$ has a temporal span of $2 \pi Q/\la$. To ensure this span is less than $T$, $\la$s are only defined for $\la
\geq 2 \pi Q/T$. For lower frequencies $[0,2 \pi Q/T]$, $Q-1$ equally spaced filters with BW $2 \pi / T$ are designed. The wavelet transform of a signal $x$ is expressed as: 
\vspace{-.1cm}
\begin{equation}
\label{eq:wavelet_transform}
	W x = \Bigl(x\star\phi(t)\,,\, x\star\psi_\la (t)
\Bigr)_{t \in \R,\la\in{\Lambda}}~.
\vspace{-.2cm}
\end{equation}
Here, $\phi$ is the low-pass filter with BW $2 \pi / T$ and the set of all higher ($\geq 2 \pi Q/T$) center frequencies are denoted by $\Lambda$. For translation invariance, as a contractive non-linear operator~\cite{mallat2012group}, modulus operation is applied over $W x$:
\vspace{-.15cm}
\begin{equation}
\label{eq:wavelet_modulus}
	|W_1| x = \Bigl(\underbrace{x\star\phi(t)}_{S_0 x(t)}\,,\, \underbrace{|x\star\psi_{\la_1} (t)|}_{U_1 x(t,\la_1)}
\Bigr)_{t \in \R,\la_1\in{\Lambda_1}}~.
\end{equation} 
\vspace{-.3cm}

The first stage of the wavelet transform applies wavelets with center frequencies $\Lambda_1$ and resolution $Q_1$. From Eq.~\ref{eq:wavelet_modulus}, we set $S_0 x(t) = x \star \phi(t)$, which is locally translation invariant due to averaging with $\phi$. The term $|x\star\psi_{\la_1} (t)|$ provides a time-frequency representation of $x$ where the varying bandwidth filters $\psi_{\la_1}$ introduce the required deformation stability~\cite{anden2014deep}. The resultant representation is then low-pass filtered with $\phi$ so as to capture long temporal context and is finally denoted as the \emph{first-order scattering coefficients} $S_1 x(t,\lau) = |x\star\psi_{\la_1}|\star\phi(t)$.

$S_1 x(t,\la_1)$, also known as \emph{scalogram}, approximates mel-spectrograms if $Q_1 = 8$~\cite{anden2014deep}. The information lost in the scalogram due to low-pass filtering is further retrieved by applying the second stage of wavelet filters:
\vspace{-.1cm}
\begin{equation}
\label{eq:2nd_wavlet}
|W_2|\, |x\star\psi_{\la_1}| = 
\Big(\underbrace{|x\star\psi_{\la_1}|\star\phi}_{S_1 x(t,\la_1)}
\,,\, \underbrace{||x\star\psi_{\la_1}| \star \psi_\lad |}_{U_2 x(t,\la_1,\la_2)}\Big)_{\la_2 \in \Lambda_2}~.
%\vspace{-.25cm}
\end{equation}
From Eq.~\ref{eq:2nd_wavlet}, we compute the \emph{second-order scattering coefficients} (also referred as modulation spectrum) $S_2 x(t,\la_1,\lad) = ||x\star\psi_{\la_1}| \star \psi_\lad| \star\phi(t) = U_2 x(t,\la_1,\la_2) \star\phi(t).$
 
The second layer coefficients capture the information lost in first layer over longer temporal context of $\phi$. Usually, we follow $Q_2 = 1$ to capture the short-spanned transients and to generate sparse representation~\cite{anden2014deep}. Iteratively, we can apply successive stages of modulus wavelet transform followed by low-pass filtering to generate the higher-order scattering coefficients. For decorrelation and to create invariance against multiplicative factors, at each layer, scattering coefficients are log-normalized by the previous layer's coefficients~\cite{anden2014deep}. Algorithm~\ref{algo:1} summarizes the process of $m$-th order WST feature extraction.

\subsection{WST for frequency transposition invariance}
The WST can further be operated along the log-frequency ($log \la$) axis of the scalograms for frequency transposition invariance.
Frequency transposition is characterized by the inter-speaker translation of spectral components in the log-frequency scale, affecting the pitch and spectral envelope information~\cite{anden2014deep}. The frequency-domain WST coefficient computation is similar to time-domain WST, with $t$ replaced by $log \la$. For LID tasks, where robustness against speaker variability is important, the exploration of frequency-domain scattering is interesting.

\begin{algorithm}[!t]
\footnotesize
\ra{1}
\caption{Computing WST features up to order $m > 1$.}\label{algo:1}
\hspace*{\algorithmicindent} \textbf{Input}: $x(t),~Q_1,~Q_2, \cdots, Q_m,~\phi$\\
 \hspace*{\algorithmicindent} \textbf{Output}:~
 $\tS x$ 
\begin{algorithmic}[1]
\Procedure{}{Applying cascaded modulus wavelet transform $|W|$ followed by averaging with $\phi$ } 
\State {$U_0 x = x$}
\For {$l=1:m$}
\State $|W_{l+1}| \,U_{l} x = (S_{l} x \,,\,U_{l+1} x)$ %\Comment{$l$-th stage wavelet transform}
\If {$l == 1$}
\State{$\tS_1 x(t,\la_1) = log \left(\frac{S_1x(t,\la_1)}{|x|\star\phi(t) + \epsilon}\right)$} \Comment{Log-normalization}
\Else
\State{$\tS_l x(t,\la_1, \cdots ,\la_l) 
= log \left(\frac{S_l x(t,\la_1, \cdots ,\la_l)}{S_{l-1} x(t,\la_1,  \cdots ,\la_{l-1})+\epsilon}\right)$}
\EndIf
\EndFor
\State {$\tS x = \left(\tS_0 x(t)\, , \, \tS_1 x(t,\la_1)\,, \cdots\,,  \,\tS_m x(t,\la_1,\la_2, \cdots, \la_m) \right)$}
\EndProcedure
\end{algorithmic}
\end{algorithm}
\setlength{\textfloatsep}{0pt}% Remove \textfloatsep
%\vspace{\baselineskip}

\section{Experiment details}
\label{sec:3}
\subsection{Database description}
We use three most widely used South Asian LID corpora, IIITH-ILSC~\cite{vuddagiri2018iiith} (IIITH), LDC 2017S14~\cite{LDC} (LDC), and IITKGP-MLILSC~\cite{maity2012iitkgp} (KGP) with five languages, \emph{Bengali}, \emph{Hindi}, \emph{Punjabi}, \emph{Tamil}, and \emph{Urdu}. IIITH and KGP data are pre-partitioned in speaker disjoint train and test sets. We manually split the LDC data into speaker disjoint train and test sets following $80:20$ ratio. The three train sets are further split into speaker disjoint train and validation sets using the same ratio. The train set is further augmented and then sampled to make it two folds following the augmentation procedure followed in~\cite{snyder2018spoken}, which applied babble, music, noise samples, and different room impulse responses as perturbations to the utterances. All the utterances are re-sampled to \SI{8}{\kilo\hertz} and split into \SI{3}{\second} chunks. With the training and validation sets of each corpus, we train three standalone LID systems. During the evaluation, with the test sets of all three corpora, we perform same-corpora and cross-corpora evaluations. We also use subset of the VoxLingua107 corpus~\cite{valk2021voxlingua107} as a blind evaluation set, entirely unknown during the system development stages.

\subsection{Data pre-processing and feature extraction}

The audio signals are first processed with an energy-based voice activity detector (VAD) to discard the silence-detected frames. Then, we extract WST features with different hyper-parameter sets, which include the temporal span ($T$) of $\phi$, the order of WST coefficients ($m$) indicating the number of layers, and octave resolutions ($Q = [Q_1, Q_2, \cdots, Q_m]$) at each layer. We use Morlet wavelet and explore it for $T=[256, 512, \cdots, 16384]$. With the sampling rate of \SI{8}{\kilo\hertz}, it approximately covers the temporal span from \SI{30}{\milli\second} to \SI{2}{\second}. However, for all three corpora, we found a prominent drop in LID performance for $T > 2048$. Hence, we consider LID performances up to $T=2048$. For this range of $T$, the signal energy contained by the $3$-rd WST layer onwards becomes gradually negligible ($<1\%$). Hence, following~\cite{anden2014deep}, we set $m=2$. For $Q_1$, we use values 2, 4, and 8. Following the literature~\cite{anden2014deep,joy2020deep,singh2021deep}, to capture the finer temporal transients, we set $Q_2 = 1$. Similarly, for the frequency-domain WST, following the literature conventions~\cite{anden2014deep,singh2021deep}, we set $m=1$ and use octave resolutions ($Q_f$) between 3 to 8. Following Algorithm~\ref{algo:1}, the time-domain WST features from the $0$-th, $1$-st, and $2$-nd layers are concatenated after log-normalization. The frequency-domain WST features are appended with the time-domain WST features for LID training and are finally processed with cepstral mean subtraction (CMS). Following the South Asian LID literature~\cite{mounika2016investigation,mandava2019investigation,dey2021cross,dey2023cross} as baseline reference, we also train the LID systems using 20-dimensional MFCCs with \SI{25}{\milli\second} window, \SI{10}{\milli\second} hop-length, 20 mel-filters, and processed with CMS. Following NIST LRE and OLR challenge protocols~\cite{sadjadi20182017,li2020ap20}, we use equal error rate (EER) and $C_{\mathrm{avg}}$ as performance metrics.

\subsection{Classifier description}
From a computation perspective, WST is similar to the convolutional neural network (CNN) architecture, while the filters are pre-defined, not learned~\cite{mallat2012group}. Hence, after the hand-crafted convolution layers, we use time-delay neural network (TDNN) based stacks of dilated CNN layers. We use the ECAPA-TDNN~\cite{desplanques2020ecapa} architecture to train the LID models\footnote{\url{https://github.com/Snowdar/asv-subtools}}. ECAPA-TDNN extends the x-vector architecture~\cite{snyder2018spoken} by replacing the frame-level TDNN layers with squeeze-excitation-based residual blocks (SE-Res2), multi-layer feature aggregation, and a channel attentive pooling layer. In different speech processing tasks, this architecture is reported to outperform several other TDNN-based models~\cite{dey2023cross,desplanques2020ecapa,kumawat2021applying}. The classifiers are trained end-to-end using 30 epochs and batch size 64. AdamW optimizer is used with additive margin (AM) softmax loss~\cite{wang2018additive}. The learning rate (LR) is $0.001$ following a reduce-on-plateau based LR scheduler with patience of 5 
 and scale $0.1$.

\begin{table*}[!ht]
\centering
%\vspace{-.25cm}
\caption{Impact of different WST hyper-parameters on LID performances (EER~(\%)~/~$C_{\mathrm{avg}}*100$) using three LID corpora.}
\vspace{-.15cm}
\label{tab:1}
\resizebox{\linewidth}{!}{%
\begin{tblr}{
  column{odd} = {c},
  column{2} = {c},
  column{4} = {c},
  column{8} = {c},
  column{12} = {c},
  column{16} = {c},
  cell{1}{1} = {r=2}{},
  cell{1}{2} = {r=2}{},
  cell{1}{3} = {c=3}{},
  cell{1}{7} = {c=3}{},
  cell{1}{11} = {c=3}{},
  cell{1}{15} = {c=3}{},
  hline{1,6} = {-}{0.08em},
  hline{2} = {3-5,7-9,11-13,15-17}{0.03em},
  hline{3} = {-}{0.05em},
}
Corpus & {Baseline\\MFCC} & T = 256 &  &  &  & T = 512 &  &  &  & T = 1024 &  &  &  & T = 2048 &  & \\
 &  & Q=2 & Q=4 & Q=8 &  & Q=2 & Q=4 & Q=8 &  & Q=2 & Q=4 & Q=8 &  & Q=2 & Q=4 & Q=8\\
IIITH & 9.74 / 11.51 & \textbf{7.53} / 8.20 & 7.56 / \textbf{8.14} & 7.75 / 8.87 &  & 8.35 / 9.71 & 7.92 / 9.11 & 10.73 / 11.41 &  & 12.58 / 14.16 & 12.69 / 13.96 & 12.58 / 14.16 &  & 15.25 / 15.79 & 19.16 / 20.32 & 20.66 / 20.72\\
LDC & 21.86 / 25.70 & 17.10 / 19.99 & 17.64 / 20.41 & 18.70 / 21.88 &  & 13.24 / 15.80 & 14.38 / 16.53 & 13.86 / 16.63 &  & \textbf{12.87} / \textbf{15.41} & 14.61 / 17.74 & 13.53 / 16.09 &  & 14.99 / 16.51 & 15.59 / 17.73 & 15.48 / 18.57\\
KGP & 12.36 / 8.75 & \textbf{5.55} / \textbf{5.83} & 8.21 / 7.51 & 8.10 / 7.91 & & 8.00 / 7.75 & 8.12 / 8.12 & 7.00 / 6.62 & & 9.75 / 9.00 & 13.00 / 13.00 & 9.00 / 8.62 & & 19.00 / 18.37 & 15.12 / 15.12 & 22.00 / 21.62 
\end{tblr}
}
\vspace{-.3cm}
\end{table*}

\begin{table*}[!ht]
\centering
% \vspace{-.2cm}
\caption{Cross-corpora LID performances (EER~(\%)~/~$C_{\mathrm{avg}}*100$) using IIITH, LDC, and KGP with different WST hyper-parameters.}
\vspace{-.15cm}
\label{tab:2}
\resizebox{\linewidth}{!}{%
\begin{tabular}{cccccccccccccccccc} 
\toprule
\multirow{2}{*}{\begin{tabular}[c]{@{}c@{}}Train\\corpus\end{tabular}} & \multirow{2}{*}{\begin{tabular}[c]{@{}c@{}}Test\\corpus\end{tabular}} & \multirow{2}{*}{\begin{tabular}[c]{@{}c@{}}Baseline\\MFCC\end{tabular}} & \multicolumn{3}{c}{T = 256} &  & \multicolumn{3}{c}{T = 512} &  & \multicolumn{3}{c}{T = 1024} &  & \multicolumn{3}{c}{T = 2048} \\ 
\cmidrule{4-6}\cmidrule{8-10}\cmidrule{12-14}\cmidrule{16-18}
 &  &  & Q=2 & Q=4 & Q=8 &  & Q=2 & Q=4 & Q=8 &  & Q=2 & Q=4 & Q=8 &  & Q=2 & Q=4 & Q=8 \\ 
\midrule
\multirow{2}{*}{IIITH} & LDC & 42.43 / 44.82 & 41.72 / 39.26 & 43.73 / 41.91 & 43.91 / 41.51 & & \textbf{35.17} / \textbf{36.77} & 37.94 / 40.25 & 39.47 / 42.78 & & 40.29 / 43.74 & 35.50 / 40.12 & 40.29 / 43.74 & & 36.30 / 39.64 & 38.13 / 43.94 & 38.16 / 43.07 \\
 & KGP & \textbf{34.83} / \textbf{32.62} & 36.11 / 33.50 & 43.51 / 38.22 & 51.85 / 40.82 &  & 41.00 / 34.00 & 40.75 / 32.87 & 44.00 / 41.62 &  & 45.75 / 44.37 & 42.00 / 40.12 & 45.75 / 44.37 &  & 48.00 / 46.75 & 52.75 / 47.87 &  49.87 / 44.12 \\
 \\
\multirow{2}{*}{LDC} & IIITH & 46.35 / 43.21 & 38.49 / 39.10 & 33.67 / 32.86 & 37.62 / 35.05 &  & 40.72 / 37.09 & 37.45 / 35.42 & \textbf{31.86} / \textbf{31.95} &  & 38.69 / 34.99 & 37.73 / 35.61 & 40.71 / 37.16 &  & 39.68 / 37.69 & 38.12 / 36.80 & 41.42 / 40.33 \\
 & KGP & 42.95 / 39.59 & 40.74 / 43.05 & 45.37 / 44.86 & 46.29 / 42.34 &  & 40.00 / 39.00 & 37.00 / \textbf{33.12} & \textbf{36.00} / 35.50 &  & 45.00 / 42.62 & 43.00 / 39.50 & 38.00 / 39.00 &  & 48.87 / 47.25 & 49.00 / 43.75 & 44.00 / 40.12\\
 \\
\multirow{2}{*}{KGP} & IIITH & \textbf{36.52} / \textbf{33.59} & 43.12 / 46.96 & 41.70 / 40.62 & 49.13 / 44.27 &  & 41.33 / 41.95 & 44.81 / 41.14 & 45.38 / 40.39 &  & 41.00 / 40.00 & 42.85 / 39.32 & 41.92 / 41.66 &  & 41.88 / 43.18 & 47.72 / 47.48 & 46.33 / 44.88 \\
 & LDC & 47.25 / 45.49 & 51.28 / 44.61 & \textbf{44.91} / 41.14 & 47.91 / 42.58 &  & 46.33 / 41.66 & 45.30 / 41.23 & 45.33 / 41.27 &  & 46.89 / 42.43 & 46.22 / 41.11 & 45.03 / 41.51 &  & 45.09 / 41.15 & 46.34 / \textbf{40.70} & 48.45 / 42.19 \\
\bottomrule
\end{tabular}
}
\vspace{-.4cm}
\end{table*}

\section{Results \& discussions}

\subsection{WST hyper-parameters and LID performances}
\begin{figure}[!ht]
  \centering
  \includegraphics[trim=0cm 0cm 0cm 
    0cm,clip,width=\linewidth]{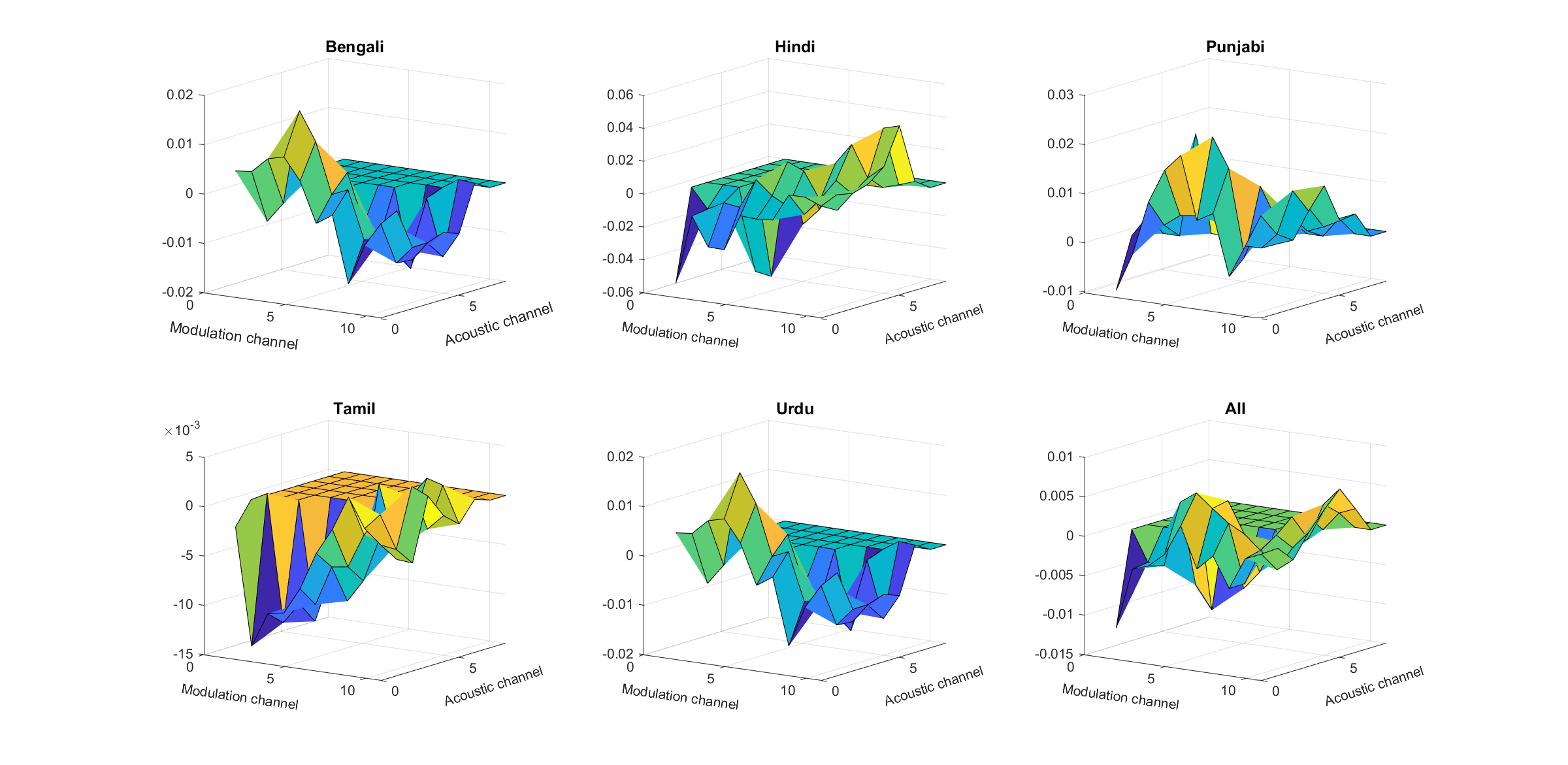}
    \vspace{-.45cm}
  \caption{Visualization of modulation spectrum for each acoustic channel, averaged over the IIITH training utterances.}
  \label{fig:scat3D}
  \vspace{-.425cm}
\end{figure}

We first extract WST features with different hyper-parameter sets by varying $T$ and $Q_1$ (as mentioned in Section~\ref{sec:3}) and train LID systems using the training and validation data of each corpus. The corresponding same-corpora evaluation performances are presented in Table~\ref{tab:1}. We use EER values to denote the best LID performances, which are used to find out the best hyper-parameter set for each corpus. The best performing hyper-parameter set for IIITH corpus ($I_{hp}$) is for $T=256$ and $Q_1 = 2$ (denoted as $Q$ in Table~\ref{tab:1}). Similarly, for LDC we obtain $L_{hp}$ for $T=1024$ and $Q=2$. For the KGP corpus, the optimal hyper-parameter ($K_{hp}$) is the same as $I_{hp}$. Two key observations from Table~\ref{tab:1} are: (i) all three corpora show the best LID performance for $Q=2$, indicating that highly localized frequency cues are not very crucial for LID tasks. (ii) IIITH and KGP both contain broadcast news reads and has optimal $T = 256$. Whereas LDC contains conversational telephone speech (CTS) and has a higher optimal $T=1024$. This observation indicates the requirement for a larger temporal context for LID in spontaneous conversations. To illustrate how the WST information is useful, in Fig.~\ref{fig:scat3D}, we plot the modulation spectrums averaged across the IIITH training utterances. The plots show distinct 3D surface patterns for different languages, indicating their efficacy in the LID tasks.

\subsection{Impact of frequency-domain WST}
\begin{figure}[!htbp]
%\vspace{-.2cm}
  \centering
  \includegraphics[width=.75\linewidth]{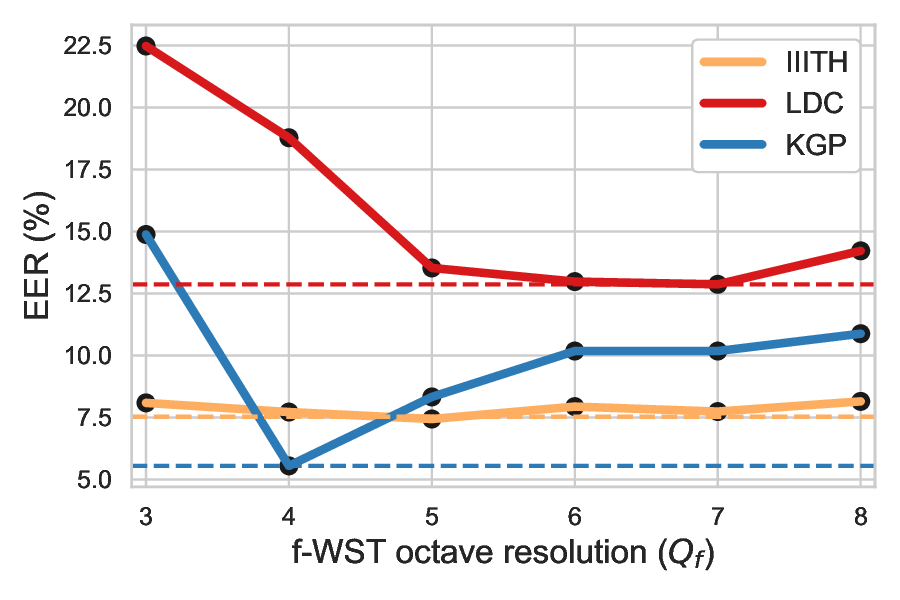}
  \vspace{-.2cm}
  \caption{Impact of f-WST octave resolution ($Q_f$) in LID performances (EER$(\%)$). Corresponding best performing time-domain WST EERs are denoted by dotted lines.}
  \label{fig:f-WST_LID}
  \vspace{-.01cm}
\end{figure}

For each corpus, we extend the best-performing time-domain WST hyper-parameters by using frequency scattering (f-WST) with $Q_f$ varying between 3 to 8. With the f-WST feature, we train LID systems and report their performances in Fig.~\ref{fig:f-WST_LID}. For comparison, we also present the best-performing time-domain WST system's LID performances and show that f-WST does not improve performance. Hence, we only consider the time-domain WST features in the subsequent experiments. We also find that KGP corpus, with the lowest number of speakers  among the three corpora, exhibits the highest EER improvement by varying $Q_f$ due to invariance for speaker variations by f-WST. While IIITH, already having much larger speakers, is inherently speaker-robust, and f-WST does not help much.

\vspace{-.2cm}
\subsection{Cross-corpora LID evaluation with optimized WST hyper-parameters}
\vspace{-.2cm}
The cross-corpora performances for the three used corpora are presented in Table~\ref{tab:2}. As reported in different literature~\cite{dey2023cross,schuller2010cross,chettri2021data}, we observe a prominent performance mismatch between the same-corpora and cross-corpora LID performances. However, we observe some improvements in cross-corpora LID performances with different WST hyper-parameters. Assume, for the train-test pair of corpus $M$ and $N$ ($M \neq N$), the best-reported cross-corpora LID performance is associated with WST hyper-parameter $(M{\text -}N)_{hp}$. Following this notation, in Table~\ref{tab:2}, we observe that $(M{\text -}N)_{hp} \neq (N{\text -}M)_{hp}$ for all three corpora. We also observe that $(M{\text -}N)_{hp} \neq N_{hp}$. Hence, the choice of optimal WST hyper-parameters depends on both train and test data. Hence, for deployment in unknown real-world scenarios, fusion of LID systems trained with WST extracted from different hyper-parameters are required. 

\subsection{Multi-WST LID system and blind evaluation}

\begin{table}[!ht]
\centering
\caption{Blind evaluation (expressed as EER~(\%)~/~$C_{\mathrm{avg}}*100$) on VoxLingua107 for the multi-WST fused LID systems.}
\label{tab:3}
\vspace{-.12cm}
\resizebox{\linewidth}{!}{%
\begin{tabular}{cccccc} 
\toprule
\begin{tabular}[c]{@{}c@{}}Training\\corpus\end{tabular} & \begin{tabular}[c]{@{}c@{}}Evaluation\\corpus\end{tabular} & \begin{tabular}[c]{@{}c@{}}Baseline\\MFCC\end{tabular} & \begin{tabular}[c]{@{}c@{}}Best\\hp\end{tabular} & \begin{tabular}[c]{@{}c@{}}Top-3\\hp\end{tabular} & \begin{tabular}[c]{@{}c@{}}Top-5\\hp\end{tabular} \\ 
\midrule
\multirow{3}{*}{IIITH} & IIITH & 9.74 / 11.51 & 7.53 / 8.20 & \textbf{4.55} / \textbf{5.80} & 4.89 / 6.10 \\
 & VoxLingua107 & 36.20 / 35.59 & 36.28 / 34.60 & \textbf{32.88} / \textbf{32.07} & 32.92 / \textbf{32.07} \\
 &  &  &  &  &  \\
\multirow{3}{*}{LDC} & LDC & 21.86 / 25.70 & 12.87 / 15.41 & 8.65 / 10.20 & \textbf{7.81} / \textbf{10.00} \\
 & VoxLingua107 & 44.28 / 43.58 & 38.56 / 38.60 & 38.31 / 38.30 & \textbf{37.88} / \textbf{37.70} \\
 &  &  &  &  &  \\
\multirow{2}{*}{KGP} & KGP & 12.36 / 8.75 & 5.55 / 5.83 & \textbf{4.87} / \textbf{4.20} & 5.15 / 4.70 \\
 & VoxLinua107 & 42.72 / 42.56 & \textbf{40.55} / 40.00 & 41.96 / \textbf{39.80} & 41.36 / 40.20\\
 \bottomrule
\end{tabular}
}
\vspace{-.01cm}
\end{table}

The optimal hyper-parameter sets for each corpus are decided based on their same-corpora LID performances. To eliminate any human-in-loop intervention in the final assessment, we conduct a blind evaluation approach using VoxLingua107~\cite{valk2021voxlingua107} corpus, following \textcolor{black}{our earlier work}~\cite{dey2023cross}. We randomly select 500 utterances from each of the five languages. Repeating it four times, we create four blind test sets, each with 500 utterances. The average LID performances over all the blind test sets are reported in Table~\ref{tab:3}. For each training corpus, we consider the top-3 and top-5 WST hyper-parameter sets and fuse the corresponding LID systems. We use logistic regression based score fusion,~\footnote{\url{https://gitlab.eurecom.fr/nautsch/pybosaris}} which are calibrated and trained with the validation set logits. The top-3 fusion systems prominently outperform the best-hp LID performance for IIITH and KGP. For LDC, top-5 fusion yields the best LID performance. From the MFCC baseline, the blind evaluation EER is decreased by $3.32\%$, $6.40\%$, and $2.17\%$, respectively, for the three training corpora.

\section{Conclusions}

To improve low-resourced LID generalization, we investigate wavelet scattering transform (WST) as an alternate feature representation. WST restores the high-frequency cues, which are lost in MFCCs with higher temporal context, as modulation spectrums. To our knowledge, this is the first work that explores WST for LID tasks. Experiments on multiple corpora show that LID tasks benefit the most with lower octave resolution in the first scattering layer. For news-read speech, smaller temporal context is desired, while the reverse holds true for conversational speech. We also find that frequency domain scattering is not beneficial for LID. Further, our cross-corpora experiments show that the optimal set of WST hyper-parameters is corpus-specific. Hence, we develop multi-WST fused LID systems for evaluation in unknown real-world scenarios. Compared to the MFCC-based baseline, the proposed system improves EER up to $14.05\%$ and $6.40\%$ for the same-corpora and blind cross-corpora evaluations, respectively. In future, we aim to develop an adaptive system to automatically obtain the optimal WST hyper-parameters depending on the data characteristics.

\bibliographystyle{IEEEtran}
\bibliography{main}

\end{document}